\documentclass[english,5p,sort&compress]{elsarticle}
\usepackage[T1]{fontenc}
\usepackage[latin9]{inputenc}
\usepackage{amsmath}
\usepackage{amssymb}
\usepackage{graphicx}
\usepackage[USenglish]{babel}
\usepackage[thinspace,thinqspace,amssymb]{SIunits}
\usepackage{color}
\usepackage{hyperref}
\begin{document}

\begin{frontmatter}{}

\title{Functional Materials for Information and Energy Technology:\\ Insights by Photoelectron Spectroscopy}

\author[pgi,jar,udu]{Martina Müller}
\author[pgi,jar]{Slavomir Nemsak}
\author[pgi,jar]{Lukasz Plucinski }
\author[pgi,jar,udu]{Claus M. Schneider}

\address[pgi]{Peter Grünberg Institut (PGI-6), Forschungszentrum Jülich, 52425
Jülich, Germany}

\address[jar]{JARA Jülich-Aachen Research Alliance, Forschungszentrum Jülich, 52425
Jülich, Germany }

\address[udu]{Fakultät für Physik, Universität Duisburg-Essen, 47048 Duisburg,
Germany}

\begin{abstract}
The evolution of both information and energy technology is intimately connected to complex condensed matter systems, the properties of which are determined by electronic and chemical interactions and processes on a broad range of length and time scales. Dedicated photoelectron spectroscopy and spectromicroscopy experiments can provide important insights. We discuss some recent methodological developments with application to relevant questions in spintronics, and towards in-operando studies of resistive switching and electrochemical processes.
 
\end{abstract}

\end{frontmatter}{}

\section{Introduction}

The development of modern technology relies strongly on advanced condensed matter systems, for example, complex layered structures. The specific functionality of a layer stack is often determined by both the layer sequence and the interfaces between adjacent layers. This holds for many systems discussed in the context of information technology, such as nano- and spin electronics, but also in devices developed for energy harvesting, conversion and storage, such as photovoltaic and fuel cells or batteries.

Common to all of the above cases is the challenge to disentangle the electronic structure and chemical states in the individual constituents and interfaces in a layer stack. Moreover, there is the desire to follow the evolution of an electronic or chemical signature \emph{in operando}, i.e. during the operation of a specific device. This may include switching processes of state variables in memory cells, charge separation in photovoltaic elements, or redox processes in electrochemical cells.

\begin{figure}
\centering
\includegraphics[width=0.75\linewidth]{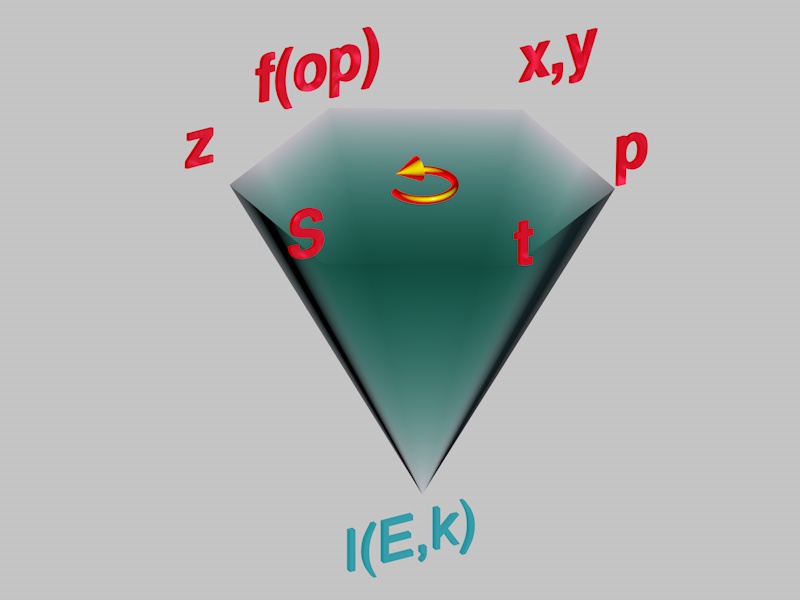}
\caption{\label{fig:Intro} Evolution of photoelectron spectroscopy from the classical surface physics tool yielding binding energy information $I(E,k)$ into a versatile method providing improved and selected probing depth ($z$), spin polarization information ($S$), or lateral resolution (${x,y}$). Current developments aim at \emph{in-operando} studies with time-resolution ($t$) while applying an external stimulus ($d$) or at near-ambient pressure ($p$).}
\end{figure}

Photoelectron spectroscopy (PES) is a very versatile tool to investigate electronic and chemical states \cite{Huf-03,RHu-05,Fad-10}. Recent years have seen several advances which further increase the capabilities of PES (Fig.~\ref{fig:Intro}). The limited information depth at soft x-ray excitation (XPS) can be overcome by exciting the photoelectrons with hard x-rays up to 10 keV \cite{Kob-09,Fad-13}. The x-ray standing wave approach provides access to specific spectroscopic information from buried interfaces \cite{FNe-14}. The evolution in spectrometer design enables XPS experiments at near ambient pressure ($\lesssim 15$ mbar) \cite{SaS-08,CBL-13} or with lateral resolution in the sub-100 nm regime \cite{EWM-05,WPK-11}. The latter spectromicroscopy concept has been successfully extended to the hard x-ray regime \cite{WTO-06,PWW-14}. 
  
In the following we will discuss some applications of advanced photoemission spectroscopy and spectromicroscopy  with soft and hard x-rays to material systems relevant for information technology and energy science.

\section{Angle-Resolved Hard X-Ray Photoemission}

Angle-resolved photoelectron spectroscopy (ARPES) has been the method of choice to investigate the electronic band structure of crystalline surfaces and novel electronic materials. Traditional ARPES employs VUV photons ($20 - 150$ eV) which in case of valence bands translates into similar kinetic energies of the emitted electrons. The inelastic mean free path (IMFP) of such electrons is $\lambda_{in}<1$~nm, translating into a surface sensitivity, which is one of the key advantages of ARPES. On the downside, however, it obscures the access to the true bulk electron dispersion and to electronic states localized at buried interfaces.

One way of increasing $\lambda_{in}$ and thus the probing depth of the ARPES experiment, is to employ low photon energies (i.e. below h$\nu$ = 10 eV). This approach, however,  is material dependent, and the kinetic energy must be in any case larger than the work function $W_F$ (for most surfaces $W_F$ is around 4 to 5 eV). Therefore, the only safe way to increase $\lambda_{in}$ is to perform HARPES (hard X-ray ARPES) experiments at photon energies in the multi-keV regime.

Since $\lambda_{in}$ should increase approximately as $E{_k}^{0.75}$, one can reach $\lambda_{in}\sim30-60$~{\AA} at 3-6~keV excitation energies. In addition a greater probing depth decreases the smearing in electron momentum perpendicular to the surface, which is proportional to $1/\lambda_{in}$ via the uncertainty principle.

\begin{figure}
\centering
\includegraphics[width=8cm]{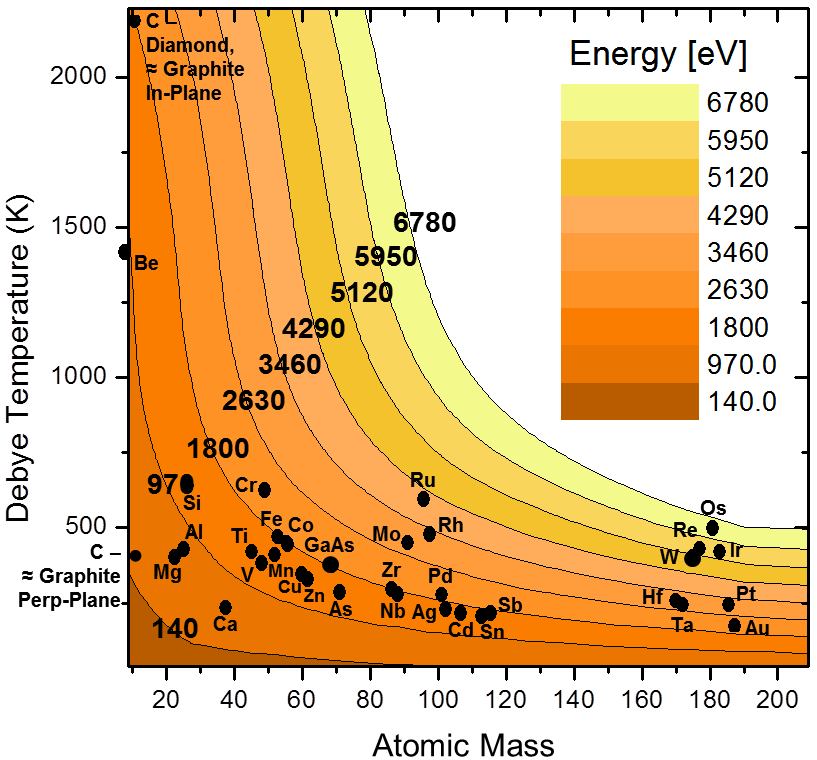}
\caption{\label{fig:HAXPES0} Contour plot showing Debye temperatures and photoelectron kinetic energies for a Debye-Waller factor W(T) = 0.5 at a sample temperature of 20~K for various elements \cite{Papp2011PRB}.}
\end{figure}

Phonon effects set a fundamental limit of the momentum resolution of the HARPES experiment. At high temperatures signatures of the band dispersions fade out in $E(\vec{k})$ maps and the valence band photocurrent reaches the matrix-element weighted density-of-states limit (MEWDOS, also often termed the XPS limit), typically also modulated by X-ray photoelectron diffraction effect \cite{Plucinski2008PRB}. Fraction of direct transitions can be estimated from the temperature-dependent Debye-Waller factor $W(T)\approx \exp{(-g^2 \langle u^2(T)\rangle)}$, where $\vec{g}$ is the reciprocal lattice vector to allow direct transitions, i.e. first Brillouin zone folding, and $\langle u^2(T)\rangle$ is the one-dimensional mean-squared vibrational displacement at temperature $T$. One can set $W = 0.5$ as a rather arbitrary limit at which realistic ARPES band mapping can be performed, and Fig. \ref{fig:HAXPES0} shows $W(T)$ contours at 20~K for selected elements \cite{Papp2011PRB} where one can see, that band mapping at up to 2~keV is possible for most elements. It turns out that this is in most cases conservative, since clear signatures of dispersions can also be clearly observed for lower $W$ values, in particular when suitable corrections for non-dispersive densities of states and photoelectron diffraction are applied to  ARPES maps of suitable signal to noise ratio. Realistic simulations of the temperature-dependent HARPES spectra have been recently performed using the one-step photoemission formalism \cite{Braun2013PRB}. Another effect which cannot be neglected at the multi-keV energy range is the photoelectron energy loss due atomic recoil when the highly-energetic electron is ejected \cite{Suga2009NJP,Takata2008PRL}.

\begin{figure*}
\centering
\includegraphics[width=14cm]{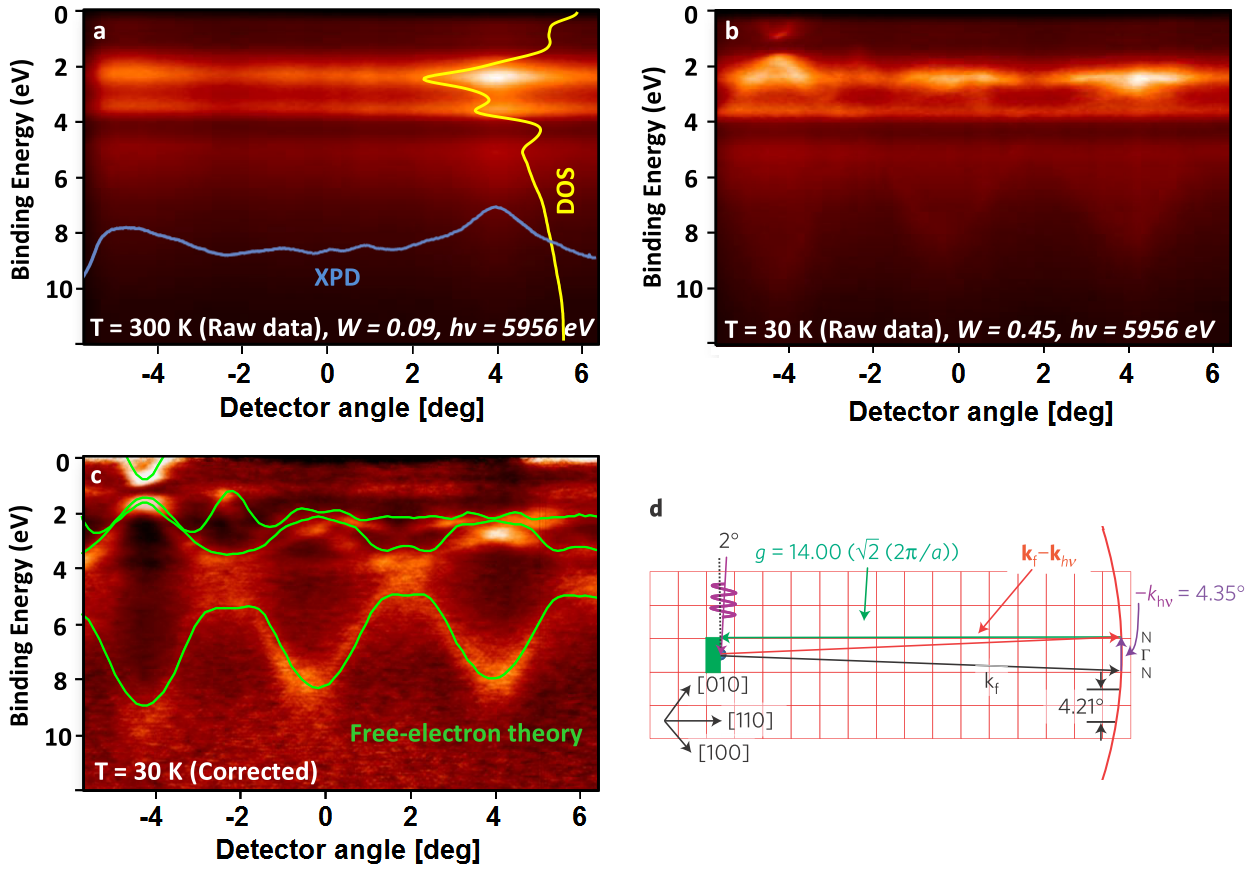}
\caption{\label{fig:HAXPES1}. Temperature dependence of HARPES $E(k)$ maps from W(110) at h$\nu$ = 6 keV \cite{Gray2011Nature}. (a) MEWDOS limit at room temperature with angular intensity modulations due to the x-ray photoelectron diffrations (XPD). (b) Clear signatures of band dispersions at T = 30~K. (c) Comparison between background-corrected E(k) map and calculated band structure.  (d) Extended BZ picture of the photoemission experiment at $h\nu$=6~keV showing the related reciprocal lattice vector $\vec G$ and the photon momentum $k_{h\nu}$.}
\end{figure*}

The effect of temperature broadening is illustrated in Fig. \ref{fig:HAXPES1}(a-b) where $E(k)$ maps measured on W(110) surface at room temperature and at 30~K are presented \cite{Gray2011Nature}. These results were obtained at h$\nu$=6~keV, which translates into 50-60~{\AA} probing depth, a true bulk sensitive band mapping.  Debye-Waller factor for tungsten at 300~K is $W = 0.09$, therefore the room temperature spectrum in Fig. \ref{fig:HAXPES1}(a) shows the MEWDOS limit. At 30~K the Debye-Waller factor $W = 0.45$ and in Fig. \ref{fig:HAXPES1}(b) one can clearly see the signatures of tungsten bulk band dispersion.

At kinetic energies above $\approx$ 500~eV the final state of photoemission experiment can be approximated by free-electron parabola according to $E_f(k_f) = \hbar^2 k_f^2/(2m_e)$, which allows for convenient interpretation of the measured valence band dispersions. Such interpretation is presented in Fig. \ref{fig:HAXPES1}(c), where free-electron final-state ground state simulations based on a state-of-the-art density functional theory (DFT) with generalized gradient approximation (GGA) are found to be in a very good agreement with measured bands. Photon momentum $k_{k\nu} = 2\pi\nu/c$, which is normally neglected in VUV ARPES, at multi-keV excitations reaches values comparable to the size of the Brillouin zone of a typical crystal. This is illustrated in Fig. \ref{fig:HAXPES1}(d), where photon momentum equals approximately the size of the entire Brillouin zone, which was taken into account in order to find the agreement shown in Fig. \ref{fig:HAXPES1}(c).

Further developments of HARPES technique will certainly involve improvements in experimental energy resolution and data acquisition times. Moreover, combining HARPES technique with the standing-wave ARPES \cite{FNe-14} will allow one to probe the electronic state localized at specific depths below the surface.

\section{Spintronics: Dilute Magnetic Semiconductors}

Recent years have witnessed a tremendous success of silicon spintronics, with considerable efforts being invested to realize applications such as e.g. a spin transistor \cite{Jansen2012}. One of the most important prerequisites for their operation is an efficient spin injection/detection into Silicon, the platform of contemporary electronics. Thereby, spin source/drain contacts need to meet two basic requirements: firstly, they must generate highly spin-polarized electron currents and, secondly, interface the semiconductor with smallest possible conductance mismatch. Various ferromagnetic $3d$ transition or half-metals have been investigated as spin-polarized source/drain materials, whereat inserting a tunnel barrier in between the silicon and metal electrode has turned out to be mandatory to ensure their spin-dependent electrical matching. Another option for the ferromagnetic electrode material comprises dilute magnetic semiconductors (DMS). However, DMS are hampered by low magnetic ordering temperatures $T_C$. A prominent dilute magnetic semiconductor is Mn-doped GaAs \cite{MOD-02}. Despite being successfully used in spin electronic devices \cite{OYB-99}, the microscopic origin of the ferromagnetism in GaMnAs is still disputed.

\begin{figure}
\centering
\includegraphics[width=\linewidth]{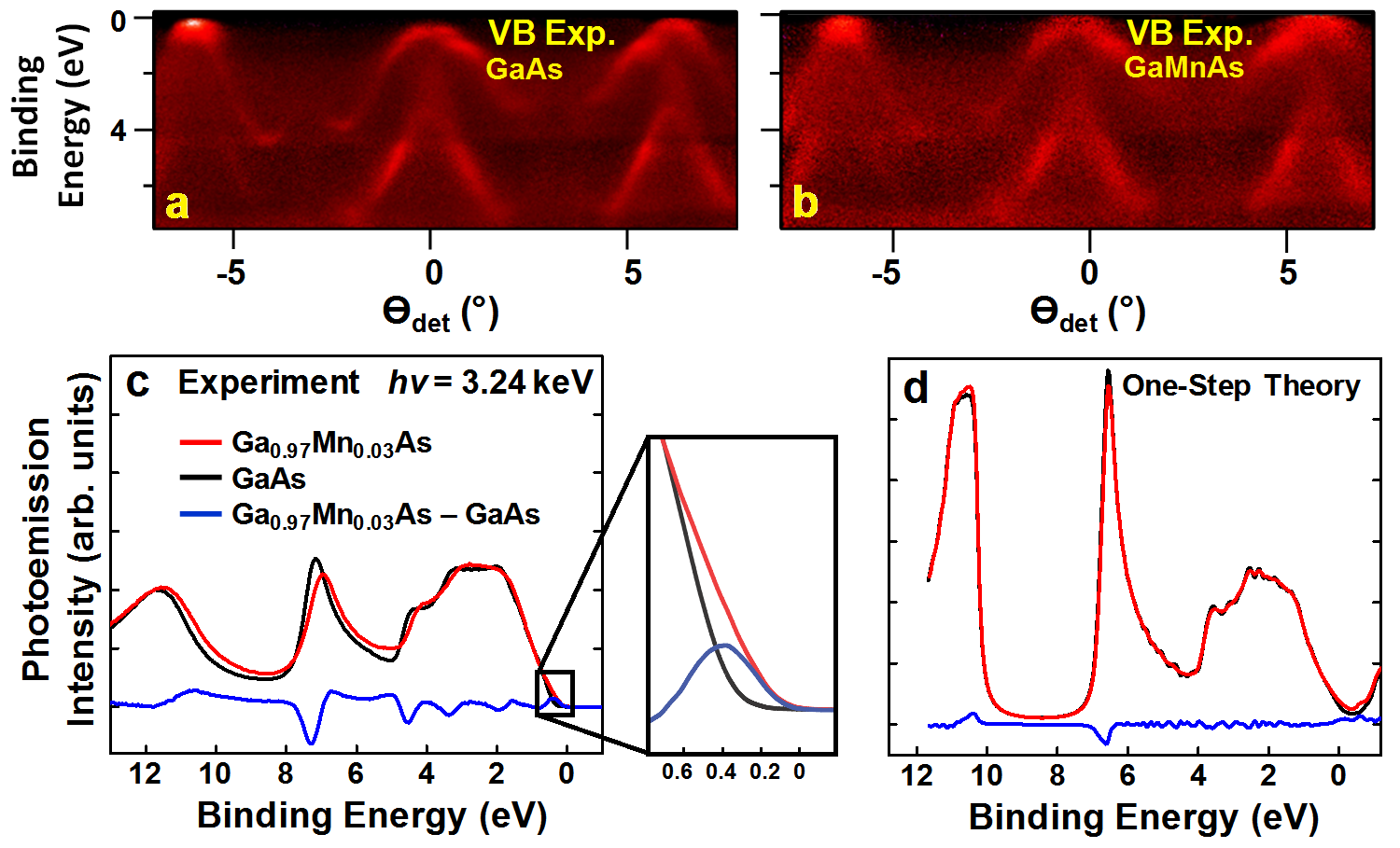}
\caption{\label{fig:HAXPES2} HARPES spectra of GaAs (a) and GaMnAs (b) measured at $h\nu$=3.24keV \cite{Gray2012Nature}. (c) Experimental angle-integrated photocurrent for GaAs (black line) and GaMnAs (red line) and the difference between them (blue line). (d) Same but for theoretical photocurrent calculated within the one-step photoemission model.}
\end{figure}

Figure \ref{fig:HAXPES2} illustrates the application of HARPES  to unravel the nature of ferromagnetism in GaMnAs \cite{Gray2012Nature}. In such a ternary disordered alloy, where some Ga sites are substituted by Mn, preparation of the stoichiometric surface is very challenging, and magnetic properties of such surface might be significantly different than bulk properties. Figure \ref{fig:HAXPES2}(a-b) shows HARPES E(k) for pure GaAs and Ga$_{0.97}$Mn$_{0.03}$As, where clear broadening of most of the spectral features is observed. However, spectra are not only broadened, but also positions and shapes of important spectral features change between GaAs and Ga$_{0.97}$Mn$_{0.03}$As, as shown in Fig. \ref{fig:HAXPES2}(c). Such semi-quantitative analysis is possible because of the increased probing depth  at high energies, which allows neglecting the spectral contribution related to the preparation-dependent surface cleanness. Figure \ref{fig:HAXPES2}(c) also shows the magnified region near the valence band maximum, which indicates the existence of the Mn-related feature 400~meV below the Fermi level in Ga$_{0.97}$Mn$_{0.03}$As, which plays a crucial role in the interpretation of origin of ferromagnetism in this material. There is no gap between this feature and the mainly GaAs-derived valence band, which suggests that magnetism originates from the coexistence of the two mechanisms -- double exchange and $p-d$ exchange, as suggested by theoretical models \cite{Sato2010RMP}. These findings are also in agreement with the theoretical one-step photoemission simulations \cite{Braun2013PRB,Gray2012Nature} shown in Fig. \ref{fig:HAXPES2}(d).

\section{ Spintronics: Magnetic Tunneling Barriers}

To date spin injection/detection in semiconductors is still inefficient, and hence a major advancement in this field is the integration of functional materials which combine the ability to generate highest spin polarization with a negligible conductivity mismatch. Magnetic oxides have recently re-emerged as highly effective spin filter materials \cite{Miao2009, Mueller2009}, and hence their integration as magnetic tunnel barriers with Silicon is of utmost interest \cite{Mueller2011}. Europium (II) oxide is widely considered as the most promising ferromagnetic barrier for silicon spintronics, because it has been shown to generate $>90\%$ current spin polarization via the spin filter tunneling effect and is predicted to be thermodynamically stable in direct contact with silicon \cite{Hubbard1996}. However, combining an ionic with a covalent material with large lattice mismatch poses a number of experimental challenges.  In particular, the control of the EuO/Silicon interface is essential to avoid the formation of metallic impurity phases, which impair an effective spin transport from the magnetic tunnel barrier into the Si electrode. Finally, stabilizing ferromagnetic, divalent Eu(II)O itself is demanding due to its metastable oxidation state in ambient atmosphere.

We succeeded in synthesizing high-quality EuO/Si heterostructures and studied the EuO ultrathin film and EuO/Si interface properties. The growth of EuO/Si heterostructure was carried out as a two-step procedure, involving an initial chemical treatment of the Si(001) surface by either wet HF etching or by a passivation of ultrathin SiO$_2$, followed by reactive Eu:O synthesis using the Eu distillation condition under UHV conditions and elevated substrate temperature \cite{Steeneken2002}. We fabricated ultrathin EuO films ($d=45$\AA) with two different compositions, i.e. stoichiometric Eu(II)O and oxygen-rich Eu(II,III), by supplying an oxygen partial pressure between $p_{O_2} =2-4 \times 10^{-9}$\,Torr. Further experimental details are given in Refs.\,\cite{Caspers2011, Caspers2013}. 

We used HAXPES to probe the electronic structure of the EuO thin films and EuO/Si interface. Experiments were carried out at the beamline P09 at PETRAIII (DESY Hamburg) \cite{Gloskovskii2011} and at the KMC-1 dipole beamline at BESSY\,II (Berlin) \cite{Gorgoi2009} with the HIKE endstation, featuring a total energy resolution of $\sim 0.5$\,eV for $h\nu=4$\,keV. 

\begin{figure*}[t]
			\centering
			 \includegraphics[width=\linewidth]{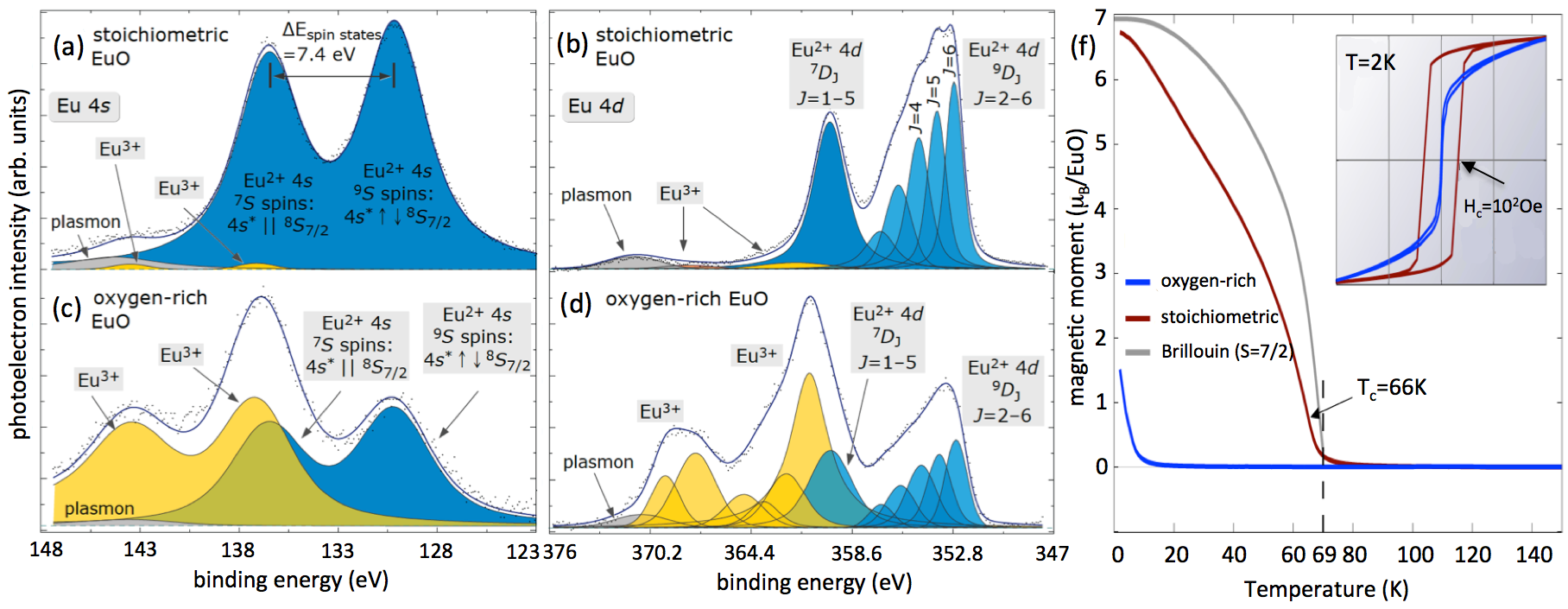}		
			\caption{(color online) HAXPES spectra of 4.5\,nm EuO directly grown on HF-etched Si(001) recorded at $h\nu=4.2$\,keV. Eu $4d$ and Eu $4s$ core levels are shown for (a),(b) stoichiometric Eu(II)O and (c),(d) oxygen-rich Eu(II,III)O films. (f) Magnetic properties of 4.5\,nm thick EuO films grown on HF-etched Si(001) dependent on temperature $M(T)$ and (inset) field-dependent hysteresis loops $M(H)$.} 
			\label{Fig:EuO1}
\end{figure*}

First, we demonstrate that stabilizing high-quality EuO directly on Silicon is feasible at optimized synthesis parameters. Fig.\,\ref{Fig:EuO1} shows the Eu $4s$ and Eu $4d$ core level spectra for (a),(b) stoichiometric Eu(II)O and (c),(d) oxygen-rich Eu(II, III). For the Eu $4s$ levels in Fig.\,\ref{Fig:EuO1}\,(a) and (c), the clear double-peak structure is caused by coupling of the $4s$ core-level electrons to the localized Eu $4f$ state, which leads to an exchange splitting $\Delta \mathrm{E}= 7.4$\,eV of the $4s$ inner shell. The spins in the $4s$ photoemission final state $4s^1 4f^7$ are predicted to antiparallel (low spin, $^7S$) for the lower binding energy peak and parallel (high spin, $^9 S$) for the higher binding energy peak with respect to the $4s$ emitted spin. For stoichiometric EuO, the $4s$ double-peak in Fig.\,\ref{Fig:EuO1}\,(a) is assigned to divalent Eu$^{2+}$ spectral contributions, whereas an additional overlapping double-peak feature appears in Fig.\,\ref{Fig:EuO1}\,(c), which is chemically shifted by $8.1$\,eV towards higher binding energy and corresponds to trivalent Eu$^{3+}$ cations. 

Due to the strong $4d-4f$ exchange interaction and much weaker $4d$ spin-orbit splitting, the prominent Eu $4d$ double peak structures in Fig.\,\ref{Fig:EuO1} (b) and (d) are assigned to a $J = L-S$ multiplet splitting, and the two peaks are denoted as $^7D_J$ and $^9D_J$ multiplets \cite{Cho1995}, respectively. The fine structure of the $^7D$ final state is not resolved, whereas the $J = 2-6$ components in the $^9D$ state are easily identified. We clearly observe a mainly divalent Eu$^{2+}$ valency in Fig.\,\ref{Fig:EuO1}\,(b), but significant spectral contributions from trivalent Eu$^{3+}$ cations in oxygen-rich EuO in Fig.\,\ref{Fig:EuO1}\,(d).

\begin{figure*}[t]
			\centering
			 \includegraphics[width=\linewidth]{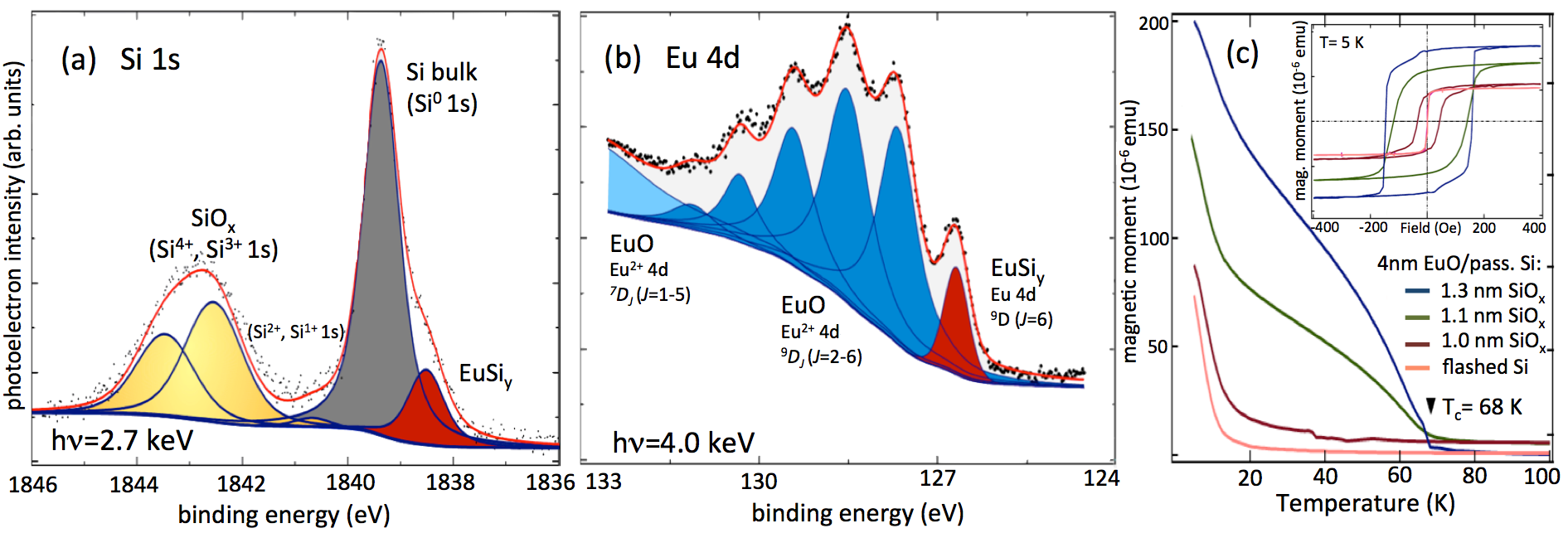}		
			\caption{(color online) (a) Si $1s$ and (b) Eu $4d$ HAXPES spectra of SiO$_x$-passivated EuO/Si(001) samples used for quantitative peak fitting analysis. (c) Magnetic properties of EuO/Si heterostructure with different SiO$_x$ passivation layers.} 
			\label{Fig:EuO2}
\end{figure*}

For both sample types, stoichiometric Eu(II)O and ox\-ygen-rich Eu(II,III)O, we performed SQUID studies (Fig.\,\ref{Fig:EuO2}(c)). In stoichiometric Eu(II)O, the $M(T)$ characteristics roughly follows a Brillouin function with $S = 7/2$. We find a Curie temperature $T_C$ of 66\,K and a magnetic saturation moment of $M_S = 6.7 \mu_B$, closely matching the bulk values of 69\,K and $7\mu_B$ per Eu$^{2+}$, respectively. The $M(H)$ curve recorded at $T=2$\,K (inset Fig.\,\ref{Fig:EuO1}(c)) displays a square-like FM hysteresis with a relatively small coercivity of $H_c=100$\,Oe. In contrast, the sample comprising oxygen-rich Eu(II,III)O reveals an almost vanishing magnetism. The magnetic saturation moment at $T=2$\,K reaches only $M_S=1.5 \mu_B$ , caused by paramagnetic Eu$^{3+}$ cations in Eu$_3$O$_4$ or Eu$_2$O$_3$ phases, which form under oxygen excess. This predominant paramagnetism is also reflected in the $M(H)$ curve (inset), in which hardly any hysteretic behavior shows up.

Noticeably, although the growth parameters for stoichiometric Eu(II)O and oxygen-rich Eu(II,III)O only slight\-ly differ in the amount supplied oxygen partial pressure by about $2 \times 10^9$\,Torr for the reactive Eu:O synthesis, the alteration of the chemical state and magnetic properties is tremendous. In that, the results underline the importance of carefully controlling the narrow parameter range during growth in order to stabilize stoichiometric -- and ferromagnetic -- Eu(II)O ultrathin films on Silicon.

Having successfully demonstrated the feasibility to synthesize ultrathin EuO tunnel barriers, in the next step we need to address the chemical state of the interface with Silicon. Special attention must be paid to the formation of metallic impurities as those can significantly decrease the spin transfer ratio across the EuO/Si interface by spin-dependent scattering processes. We therefore studied the impact of an interface passivation with the native silicon oxide SiO$_2$, in order to define in a controlled manner an additional -- but nonmagnetic -- tunnel barrier, which as such does not alter the spin transfer efficiency. 

Interface-sensitive HAXPES spectra of the EuO/Si heterostructures are depicted in Fig.\,\ref{Fig:EuO2}\,(a),(b). In the Si $1s$ spectrum (a), SiO$_x$ contributions are observed at chemical shifts of $3.2$ and $4.1$\,eV, which identify Si$^{3+}$ (Si$_2$O$_3$) and Si$^{4+}$ (SiO$_2$). Another component with a chemical shift to lower BE is identified as a metallic Eu silicide (EuSi$_y$) contribution both in the Si$1s$ and Eu $4d$ spectrum in Fig.\,\ref{Fig:EuO2}\,(a),(b). A quantitative thickness determination of the SiO$_x$ passivation layer and of the interfacial EuSi$_y$ reaction layer is accomplished by consistent least squares peak fitting \cite{Tanuma2011, Eickhoff2004, Caspers2013}. We found EuSi$_y$ to exceed 10\AA \, thickness for flashed Si(001), however, by applying an ultrathin SiO$_x$ interface passivation ($d_{\mathrm{SiO}_x}=10-13$\AA), the silicide formation is suppressed by $68\%$ to $d_{\mathrm{EuSi2}} \leq 7.5$ \AA. The optimum value for the quasi-contamination free EuO/Si interface is obtained with 13\AA  \ SiO$_x$ passivation, for which the silicide is diminished down to 1.8\AA, well below one monolayer of interfacial coverage.

Finally, we show how the magnetic properties of the EuO/Si heterostructures depend on the SiO$_x$ interface passivation. Fig.\,\ref{Fig:EuO2}\,(c) summarizes
the in-plane $M(T)$ and $M(H)$ SQUID curves with varying SiO$_x$ thicknesses. Not surprisingly, EuO thin films on Si(001) without any SiO$_x$ passivation show a significantly reduced magnetic moment and a $T_c \sim 10$\,K, as the formation of a EuSi$_y$ reaction layer largely destroys the EuO ferromagnetic behavior \cite{Caspers2013-2}, as observed also by HAXPES. Applying the SiO$_{\mathrm{x}}$ interface passivation in turn clearly improves the EuO magnetic properties. The most effective interface passivation of $13$\AA \, SiO$_x$ yields  $M_S=5 \mu_B$ and $T_C = 68$\,K, close to the respective bulk values. With higher SiO$_x$ passivation thicknesses, however, structural defects and surface roughening cause the EuO coercive field to increase up to 180 Oe.

These results demonstrate that stabilizing high quality EuO magnetic tunnel barriers directly on Silicon is feasible by carefully controlling the narrow range of synthesis parameters. Inserting an ultrathin SiO$_2$-interface passivation layer acts as a reaction barrier for the transport interface by significantly reducing silicide formation. In thus, the integration of spin filtering and conductance-matched magnetic tunnel barriers directly with Silicon may open up novel pathways for the upcoming developments in silicon spintronics.

\section{Resistive Switching: HAXPEEM}

In order to reduce the energy consumption of information technology, a fast high-density nonvolatile memory is needed. Among the different avenues followed is the concept of resistive switching in oxides \cite{WAo-07} or chalcogenides \cite{WYa07}. The physical mechanisms underlying resistive switching include redox processes and phase changes. A particular challenge in the investigation of resistive switching phenomena is the small area where the changes take place. With a lateral resolution of a few 10 nm, energy-filtered photoemission microscopes are ideally suited to tackle this challenge. Studies in the soft x-ray regime have already contributed significantly to the understanding of the microscopic processes during electroforming in Fe-doped SrTiO$_3$  \cite{LPM-14} and resistive switching in GaO$_x$ \cite{AWF-14}.

The major drawback of using soft x-rays, however, is the limited information depth of the photoemission signal. This precludes \emph{in-operando} studies of the switching process, as most resistive switching contacts involve a vertical contact geometry with a metallic top electrode. Accessing the chemical changes in the oxide chalcogenide layer therefore requires a delamination of the top electrode -- which may cause unwanted artefacts in the experiment. In order to overcome this problem we are currently developing energy-filtered hard x-ray photoemission microscopy (HAXPEEM) \cite{WPC-12,PWW-14}. Using photoelectrons with kinetic energies of around 5-6 keV permits access to interfaces buried below a 10 nm thick top electrode.

\begin{figure}
\centering
\includegraphics[width=8cm]{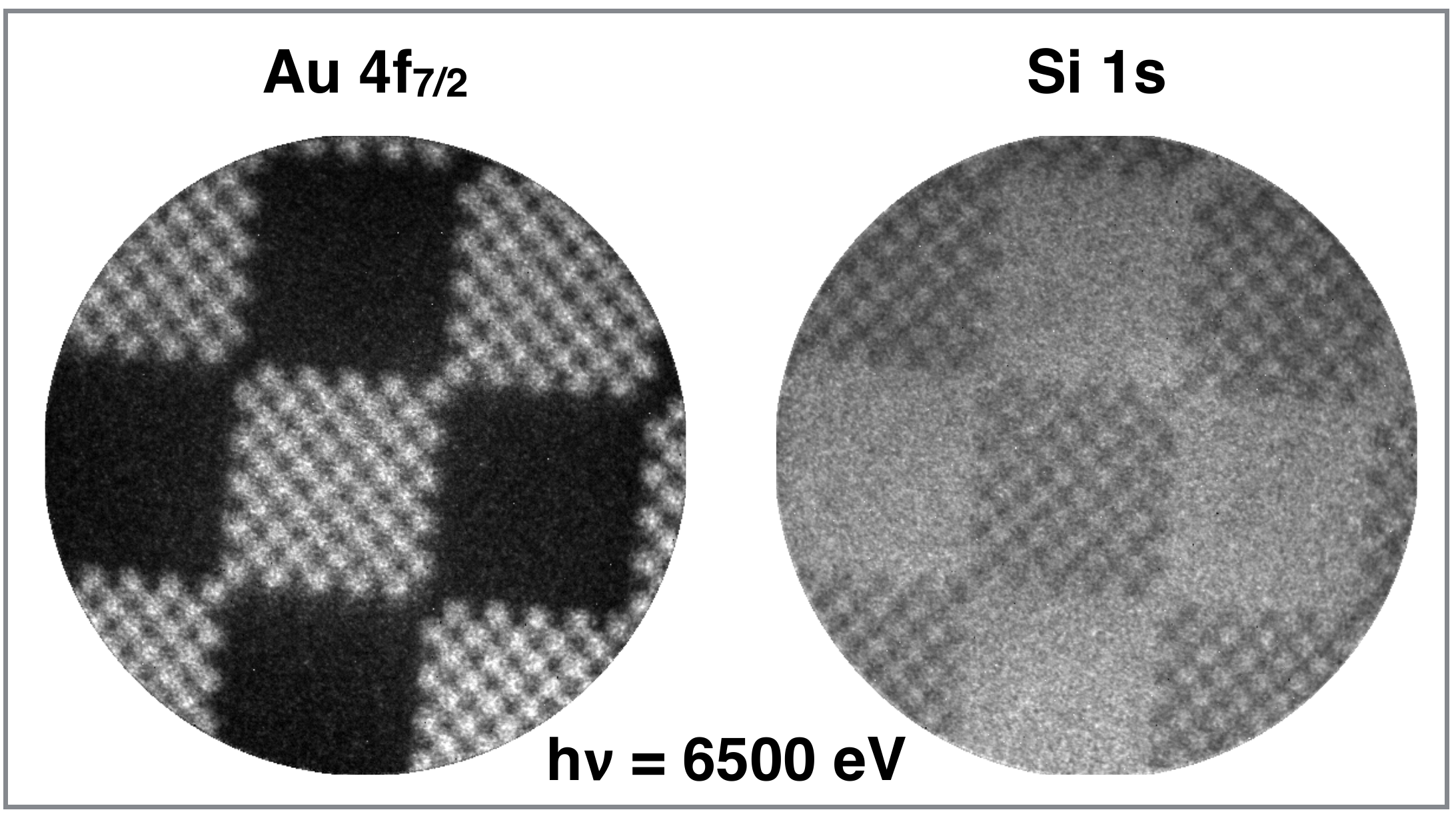}
\caption{\label{fig:HAXPEEM1} Hard x-ray spectromicroscopy from a test structure. The checkerboard pattern is formed by $1\mu$m edge length square Au elements on Si, which are grouped into $10\mu$m edge length arrays.  Au $4f_{7/2}$ (left) and Si $1s$ photoemission signal (right). After ref.~\cite{WPC-12}.}
\end{figure}

A result from a first HAXPEEM test experiment on a Au/Si checkerboard structure is shown in Fig.~\ref{fig:HAXPEEM1}. We used photons with an energy of 6.5 keV delivered by the beamline P09 at PETRA-III (Hamburg). The images were acquired with a modified NanoESCA system capable of energy filtering up to a electron kinetic energy of 6 keV \cite{WPC-12}. At the Au $4f_{7/2}$ line the kinetic energy is about $E_{kin}\sim6.4$~keV and the image shows a clear chemical contrast. The image recorded at the Si $1s$ line has an inverted contrast, but is more noisy as the respective photoemission signal is only half the size of the signal from Au. This result demonstrated that hard x-ray spectromicroscopy is possible with a lateral resolution of at least 500 nm.

\begin{figure}
\centering
\includegraphics[width=9cm]{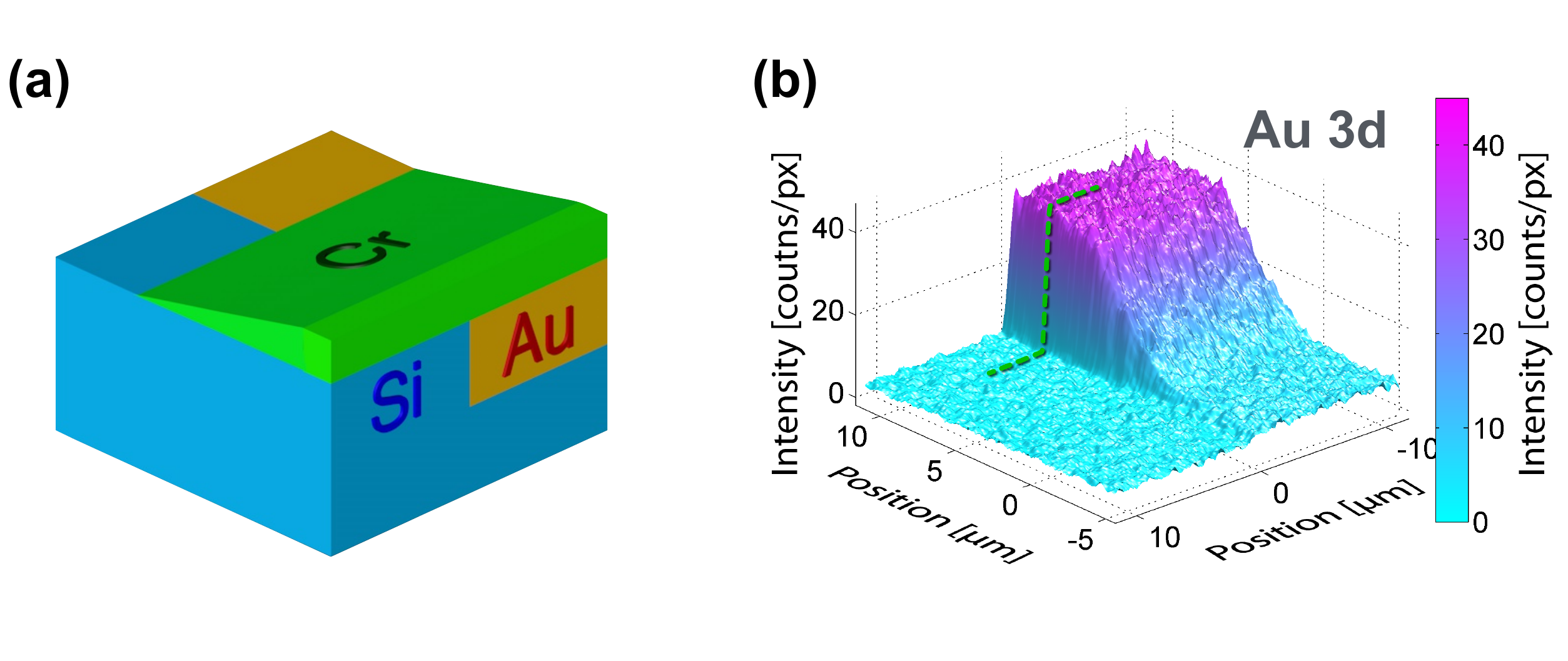}
\caption{\label{fig:HAXPEEM2} Hard x-ray spectromicroscopy from a wedge-shaped test structure. (a) Sample layout, thickness of Au inlay is 100~nm. The maximum thickness of the Cr top layer is 15~nm. (b) Lateral distribution of the Au $3d$ photoemission signal, reflecting the attenuation through the Cr overlayer. Adapted from ref.~\cite{PWW-14}.}
\end{figure}

In order to address the aspect of information depth we have designed a specific test sample with a wedge-shaped Cr top layer covering a sharp transition between Si and Au (Fig.~\ref{fig:HAXPEEM2}a). The results displayed in Fig.~\ref{fig:HAXPEEM2}b reflect the lateral distribution of the Au $3d$ photoemission signal at a photo excitation with $h\nu=6550$~eV. The sharp edge separating the Si and Au regions is clearly visible. The gradual decrease of the Au $3d$ photoemission signal reflects the attenuation of the photoelectrons with a kinetic energy of $E_{kin}=4342$~eV in the top Cr layer. From a modelling of this decrease we obtain an inelastic mean free path of $\lambda_{in}\approx 4.9$~nm \cite{PWW-14}. 

These results demonstrate that hard x-ray spectromicroscopy has reached a state that enables future in-operando experiments on resistive switching materials. 

\section{Probing Liquid Interphases}

Many areas of science and technology -- from energy generation and storage, environmental science and corrosion to electrochemistry -- specifically require studying of buried interfaces. Probing of such interfaces, either solid/liquid, solid/gas or liquid/gas, pose major experimental challenges when investigating their chemical and structural properties on the nm or sub-nm scale.  For example, the electrical double-layer has been studied for over 100 years, and yet is not fully understood \cite{Ohno2011,Brown2012}. Ambient pressure photoemission is an excellent tool for exploring liquid and gas interactions with solid in terms of chemical and charge information \cite{Bluhm2013}, but is limited by its surface sensitivity. Here presented combination of standing-wave and ambient-pressure photoemission (SWAPPS) partially suppresses this limitation and provides a direct way to study a solid/liquid and a liquid/gas interface via selectively enhancing a signal originating from different depths.

\begin{figure}
\centering
\includegraphics[width=8.5cm]{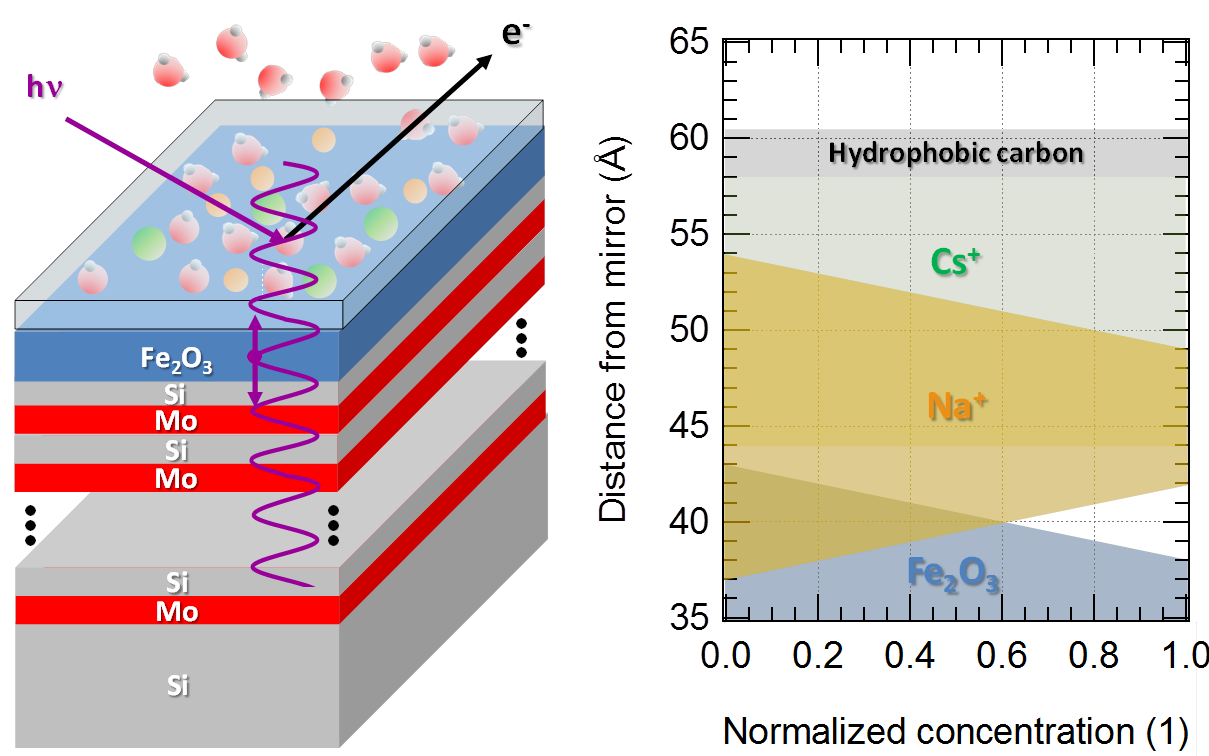}
\caption{\label{fig:APPS1} Soft X-ray standing wave study of a hematite sample hydrated in ambient water vapor pressure. Left - an experimental schematic of the multilayer mirror, the hematite layer, the wet layer and the humid environment.  Right - depth resolved concentration profiles of various chemical species.  Adapted from \cite{Nemsak2014}.}
\end{figure}

A model sample (Fig. \ref{fig:APPS1}) is represented by a hydrated, mixed NaOH + CsOH layer on a polycrystalline hematite (Fe$_2$O$_3$) thin film \cite{Nemsak2014}. The liquid film is prepared by exposing the sample to well controlled relative humidities just below saturation. The precise control of sample temperature and liquid (water) vapor pressure in the experiment is crucial, since any small deviation would lead to either a rapid liquid film growth or desorption. Another problem resulting from the liquid film thickness variations is a complicated control of the concentration and pH of the re-hydrated solution.  While alkali halide solution films have successfully been formed by re-hydrating either drop-cast or evaporated alkali halide films inside a humidity-controlled vacuum chamber \cite{Arima2010,Nemsak2014}, the exact control of the concentration and pH of the film remains challenging.

The hematite film is grown by sputtering deposition onto multi-layered mirror, which acts as a strong standing wave (SW) generator. By tuning both the photon energy and the beam incident angle to follow the Bragg condition, a SW perpendicular to the sample surface is formed. Rocking the sample around the Bragg angle then causes tailoring SW nodes and antinodes along the sample normal, resulting in the depth variable selectivity of this technique. Photoemission signal recordings from different core-level as a function of incident angle (so-called ``rocking curves'') then carry signatures of the elements' position in depth. A careful analysis of the rocking curves of all the elements of interest can then provide a detailed model for depth distributions and concentration gradients of different species \cite{Fad-13,Gray2014}.

\begin{figure*}
\centering
\includegraphics[width=14cm]{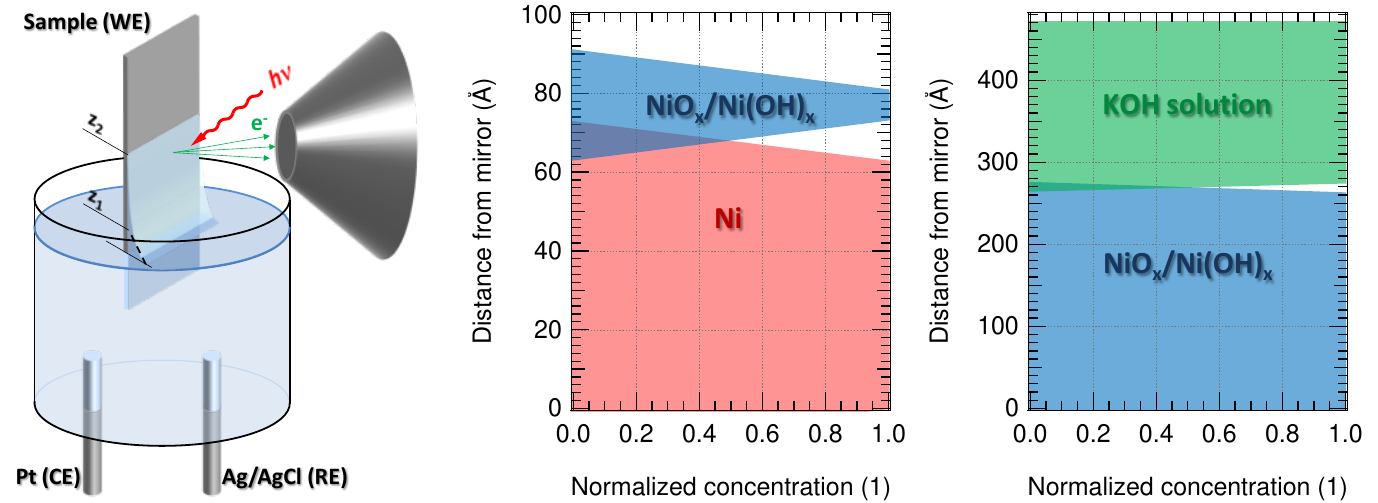}
\caption{\label{fig:APPS2}  Hard X-ray standing wave \emph{in-operando} photoemission study of nickel oxidation using a meniscus method. Left: schematics of the meniscus method. Counter, reference and working electrodes are labeled (CE, RE, WE), with the working electrode (sample) being grounded. Center and right panels: depth concentration profiles of various species for a dry and dipped sample, respectively. Adapted from \cite{KNZ-15}.}
\end{figure*}

Fig. \ref{fig:APPS1} shows the depth profiles obtained from the analysis of different spieces' rocking curves using YXRO simulation package \cite{YXRO}. As the first noteworthy result, we show that the Na$^+$ and Cs$^+$ ions have distinctly different depth distributions within the liquid layer, more specifically that Cs$^+$ ions are on average by 0.4 nm  farther away from the hematite surface than Na$^+$ ones. Also, the boundaries of the Na$^+$ presence follow the roughness of the underlying hematite, which agrees well with an assumption of the specific adsorption. On the contrary, Cs$^+$ ions keep their solvation shells complete and are not in a direct contact with hematite. Another striking result is the recognition of the hydrophilic (high binding energy) and hydrophobic (low binding energy) carbon species, which again occupy different layers within or on top of the liquid layer, respectively. Obtaining of such precise depth-resolved concentration profiles is simply not accessible by conventional ambient pressure XPS.

Another, very versatile possibility to create thin liquid films is a so-called meniscus method, which was first introduced decades ago by Bockris \cite{Bockris1969}. The principle of the method is  shown in Fig. \ref{fig:APPS2}. One of the main advantages is a suitable design for contacting the liquid layer and hence performing $in-operando$ electrochemical experiments, as has recently been successfully demonstrated for the first time to studies of liquid/solid interfaces using ambient pressure XPS \cite{Axnanda}.

The sample, a nickel sputtered layer for purposes of the SW generation again grown on a multi-layer mirror, is immersed in a 0.1M KOH solution under oxidation conditions inside the measurement cell \cite{KNZ-15}. The vacuum chamber is backfilled with a vapor close to the saturation pressure from an attached reservoir in order to keep the environment in the cell stable and to maintain the liquid film after retracting the sample from the cell.

After electrochemical treatment (cleaning/sample oxidation/reduction), the sample is partially pulled out from the solution and a meniscus extends to a height $z_1$ above the bulk solution surface.  Under favorable conditions and in the above mentioned presence of a vapor pressure (which is a function of the solution temperature) close to saturation, a thin liquid film can be stabilized over the whole originally dipped length of the sample ($z_2$).  Close to the position $z_2$), parts of the liquid film are sufficiently thin to allow collection of photoelectron spectra from the liquid/solution interface. Liquid films formed in that manner have been shown to be stable for many hours \cite{Axnanda}.

The meniscus method can provide stable layers of higher thicknesses, which is another essential property for $in-operando$ electrochemical experiments in liquids, due to mass and charge transport limitations. Liquid layers of higher thicknesses on the other hand cause problems due to the relatively small escape depth of photoelectrons \cite{Tanuma2011}, which can be partially overcome by using higher energy excitation. The combination of a standing-wave approach and hard X-ray photoemission ($h\nu=3$ keV) was used to determine depth resolved composition of the pristine and oxidized Ni layer (Fig. \ref{fig:APPS2}) \cite{KNZ-15}. Obtained concentration profiles reflect a transformation of nickel electrode from metal into oxide/hydroxide after submerging into the KOH solution. The whole volume of the nickel layer is oxidized/hydroxidized upon submerging, with solution overlayer being over 10 nm thick.

The last aspect deserving our attention are the benefits of using hard versus soft x-ray excitation in SW ambient pressure photoemission. Reflectivity, the driving force behind the SW generation, is decreasing while increasing excitation energy. Another challenge to be overcome while going from soft to hard X-rays are decreasing photoelectron cross-sections and lower incident angles needed in order to follow the Bragg law, which is problematic with a finite beam size and small field of view of ambient pressure hemispherical analyzers.

\begin{figure}
\centering
\includegraphics[width=8cm]{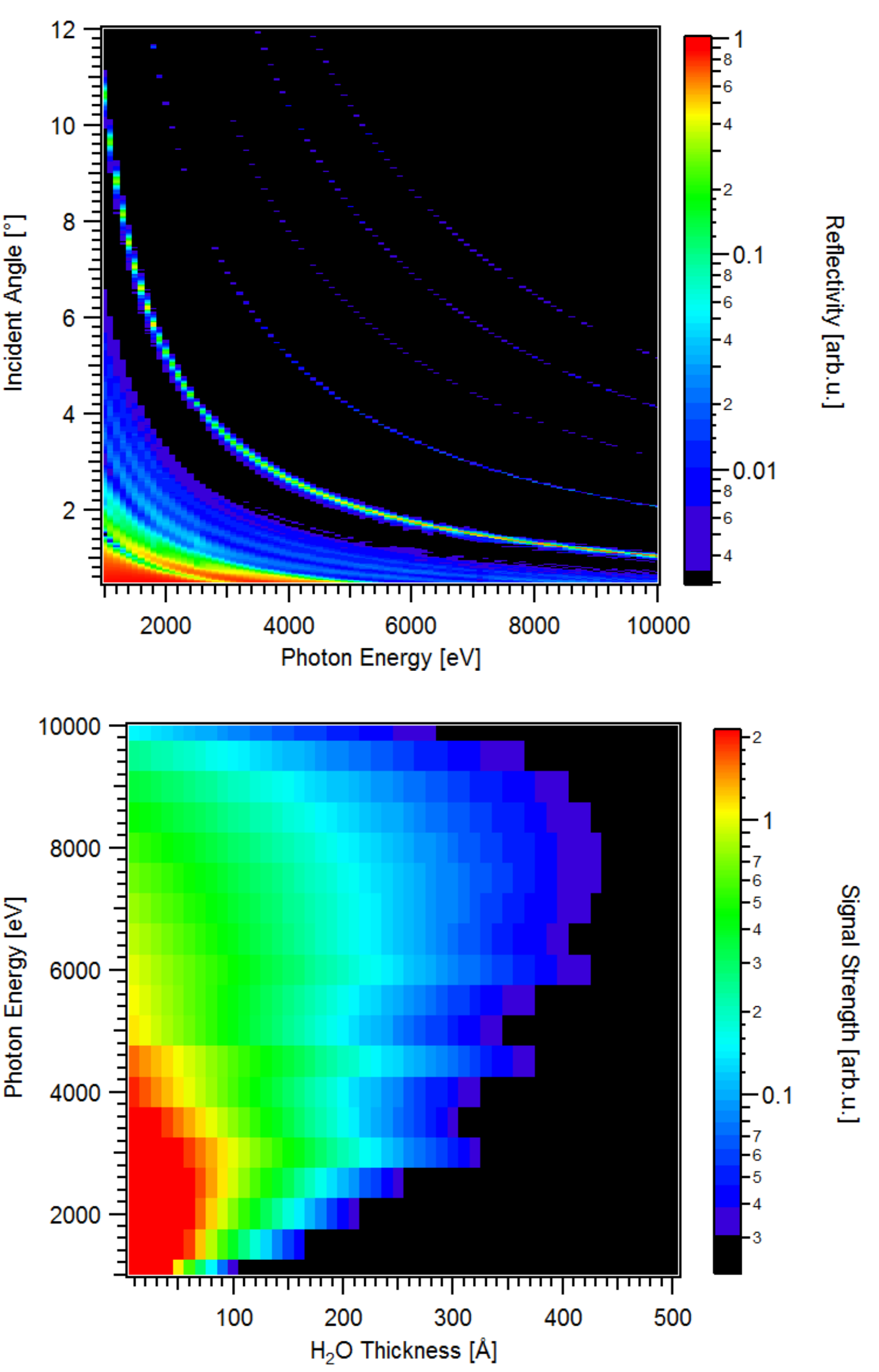}
\caption{\label{fig:APPS3}Photoemission intensity of O1s signal originating from the solid-liquid interface (0.5 nm delta layer) as a function of photon energy and liquid overlayer thickness. Beam incident angle is varied following the Bragg law.}
\end{figure}

 X-ray optical and photoemission simulations were again carried out using the YXRO software package \cite{YXRO}. A model system equivalent to the one depicted in Fig. \ref{fig:APPS1} was simulated using a variable liquid overlayer thickness and photon energy, with the beam incident angle following the Bragg peak. Photoemission signal of O 1s electrons originating from a 0.5 nm thin interfacial layer is then used as a measure of an ``interfacial signal'' strength shown in Fig. \ref{fig:APPS3}. Despite all the mentioned problems, solid-liquid interfaces can be probed by ambient pressure XPS for liquid layers of up to $\approx$ 10 nm thick when using multi keV excitation. The optimal photon energy for the SWAPPS for such thick liquid films lies in a surprisingly narrow range located between 2 and 3 keV, which overlaps with the energy range suggested by Axnanda for conventional ambient pressure photoemission experiments \cite{Axnanda}. In this energy regime and 10 nm thick overlayer, the O 1s signal intensity from the 0.5 nm interface layer is $\approx 20$ times lower comparing to bulk water.

 The combination of the ambient pressure photoemission, $in-operando$ electrochemical experiments and the standing wave approach is a novel and very promising way of studying solid/liquid interfaces relevant both in fundamental and applied research. Possibilities concerning number of systems and materials to be studied are practically unlimited.

\section{Outlook}
The examples discussed in this contribution illustrate the growing power of modern photoelectron spectroscopy to address specific issues in functional material systems. The method development in the near future clearly aims at the combination of seemingly disjunct capabilities, for example, joining lateral and temporal resolution, introducing spin polarization analysis into HAXPES, or establishing versatile pump-probe schemes to investigate processes with high time resolution.  


\section*{Acknowledgement}
We are grateful to Prof. C. S. Fadley for the ongoing collaboration on many aspects of soft and hard x-ray photoemission. We acknowledge experimental work by C. Caspers. M.\,M. acknowledges financial support by the Deutsche Forschungsgemeinschaft (DFG) under Grant MU\-3160/1-1 and by the Helmholtz Association (HGF) under Contract No. VH-NG-811. C.\,M.\,S. acknowledges support of the DFG through the CRC 917.



\bibliography{Plucinski_biblio}

\end{document}